\documentclass{article}
\usepackage{graphicx} 
\usepackage{siunitx}
\usepackage{listings}
\usepackage{hyperref}
\usepackage{nomencl} 
\makenomenclature
\usepackage{amsmath}
\usepackage{algorithm}
\usepackage{algpseudocode}
\usepackage{bm}
\usepackage{booktabs}
\usepackage[letterpaper, total={6in, 9in}]{geometry}
\usepackage{breqn}
\usepackage{tabularx}

\usepackage{multirow}
\usepackage{longtable}
\usepackage{booktabs}
\usepackage{svg}
\usepackage{wrapfig}

\title{Accurate Reduced Floating-Point Precision Implicit Monte Carlo
\footnote{
This is an Accepted Manuscript of an article published by the American Nuclear Society in Proceedings of the International Conference on Mathematics and Computational Methods Applied to Nuclear Science and Engineering (M\&C 2025) on April 27-30 2025, available at: \url{doi.org/10.13182/MC25-47193}}
\footnote{LA-UR-24-31859}}

\author{
    Simon Butson\textsuperscript{1}\footnote{Contact: butsons@oregonstate.edu}
    \and Mathew Cleveland\textsuperscript{2}
    \and Alex Long\textsuperscript{2}
    \and Todd Palmer\textsuperscript{1}
}

\date{%
    \small{
    \textsuperscript{1}School of Nuclear Science and Engineering, Oregon State University, Corvallis, Oregon; \\
    \textsuperscript{2}Los Alamos National Laboratory, Los Alamos, New Mexico
    }
}

\begin{document}

\maketitle
 
\begin{abstract}
This work describes methodologies to successfully implement the Implicit Monte Carlo (IMC) scheme for thermal radiative transfer in reduced-precision floating-point arithmetic. The methods used can be broadly categorized into scaling approaches and floating-point arithmetic manipulations. Scaling approaches entail re-scaling values to ensure computations stay within a representable range. Floating-point arithmetic manipulations involve changes to order of operations and alternative summation algorithms to minimize errors in calculations. The Implicit Monte Carlo method has nonlinear dependencies, quantities spanning many orders of magnitude, and a sensitive coupling between radiation and material energy that provide significant difficulties to accurate reduced-precision implementations. Results from reduced and higher-precision implementations of IMC solving the Su \& Olson volume source benchmark problem are compared to demonstrate the accuracy of a correctly implemented reduced-precision IMC code. We show that the scaling approaches and floating-point manipulations used in this work can produce solutions with similar accuracy using half-precision data types as compared to a standard double-precision implementation.
\end{abstract}

\section{Introduction}\label{sec:1}
With the introduction of newer computing architectures designed to support reduced-precision floating pointing arithmetic, there is greater interest in implementing reduced-precision compatible algorithms. The primary advantage of reduced-precision is decreased memory usage which could allow for bigger and faster simulations, but it can be challenging to produce accurate results for many problems including thermal radiation transport \cite{alexlong24}. The goal of this work is to explore techniques for implementing an accurate Implicit Monte Carlo (IMC) \cite{FLECK1971313} scheme for thermal radiative transfer using 16-bit (half-precision) floating-point numbers. The results of half- and double-precision implicit Monte Carlo implementations of the Su \& Olson volume source benchmark problem \cite{SU19971035} will be compared. While the strategies discussed in this paper are broadly applicable to various Monte Carlo particle transport problems, thermal radiation transport has a few qualities that make it particularly challenging to model with reduced-precision. These include various nonlinear terms, quantities that can span many orders of magnitudes and a strong time-dependent coupling between radiation and material energy. The following work was implemented in the Julia programming language \cite{bezanson2017julia}. Julia was chosen because of its native software support for half-precision floats and random number generation.  The code to produce the results in this paper can be accessed on GitHub at \url{https://github.com/simonbutson/Reduced-Precision-IMC}. 

\section{Floating-Point Arithmetic Primer}
Floating-point numbers are a way to represent subsets of the real numbers using a fixed precision. They consist of an integer significand $d$ with precision $p$ which is multiplied by an integer exponent $e$ of a given base $b$ \cite{GOLDBERGD1991Wecs}. An additional sign bit can also be included to indicate if the value is positive or negative. A floating-point number can be precisely represented in decimal notation as: 
\begin{equation}
    \pm d_0 . d_1 d_2 \space ... \space d_{p-1} \times b^e
\end{equation}
or equivalently as the sum:
\begin{equation}
    \pm (d_0 + d_1 b^{-1} + ... + d_{p-1} b^{p-1})b^e, \quad (0 \leq d_i \leq b). 
\end{equation}

In this work we will be using floating-point arithmetic as implemented by the IEEE 754 standard \cite{4610935}. Included in this standard are a few special quantities including NaN (Not a Number) and $\pm$Inf (Infinities) that are triggered in cases such as dividing by zero instead of halting the computation completely. Nonetheless, we will seek to avoid triggering them in Monte Carlo code as they can propagate  into the results of calculations in which they are involved and lead to non-physically meaningful values. 

Table \ref{tab:floating-points} below summarizes relevant parameters of 16, 32, and 64-bit floating point numbers. The range of values that can be represented using half-precision is significantly smaller than that of its higher precision counterparts. This motivates the development of some of the methods discussed in the next section. Denormalized floating-point numbers as specified by IEEE-754 are available in Julia. However, they are not considered in our work as they lack widespread hardware adoption, but they may have utility in extending the lower range of reduced-precision variables.

\begin{table}[ht]
    \centering
    \caption{Floating-Point Precision Comparison.}
    \begin{tabularx}{\textwidth}{|l|X|X|X|}
        \hline
        \textbf{Base 2 Floats} & \textbf{Half-Precision} & \textbf{Single-Precision} & \textbf{Double-Precision} \\ \hline
        \textbf{Number of Bits} & $16$ & $32$ & $64$ \\ \hline
        \textbf{Smallest Normal Positive Value} & $6.10 \times 10^{-5}$ & $1.18 \times 10^{-38}$ & $2.23\times 10^{-308}$ \\ \hline
        \textbf{Largest Value} & $65504$ & $3.40 \times 10^{38}$ & $1.80\times 10^{308}$ \\ \hline
        \textbf{Significand Digits (Binary/Decimal)} & $11 / 3.31$ & $24 / 7.22$ & $53 / 15.95$  \\ \hline
    \end{tabularx}

    \label{tab:floating-points}
\end{table}

\section{Methods}
This section will outline changes that are necessary to implement Monte Carlo algorithms in reduced-precision. These methods can be split into two broad categories: scaling approaches and floating-point arithmetic manipulations. Some additional remarks regarding reduced-precision random number generation and the batching of particles are also included.

\subsection{Scaling Approaches}
We begin by looking at scaling approaches that can be applied to reduced-precision Monte Carlo codes. These scaling approaches have two primary aims, the first being to keep the value of calculated quantities within the representable range of their respective data type. The second aim is to optimize the scale of values in order to minimize round-off errors that occur from calculations.

For starters, it is advantageous to use relative positions where distances are tracked merely within the width of a single cell and a separate cell index variable keeps track of absolute position. In a 1-D grid of uniform cell width $dx$ an absolute position $x_{abs}$ can be represented by a relative position $x_{rel} = x_{abs} - n*dx$ where $n = \left \lfloor{\frac{x_{abs}}{dx}}\right \rfloor$ is the cell index. Relative positions can be extended to non-uniform and/or higher-dimensional grids by the use of arrays and index variables.

The first advantage is to ensure consistency in distance calculations and to better evaluate when boundaries are crossed. Given the limited number of significant digits (mantissa bits) in reduced-precision, the set of representable values is not uniformly spaced throughout its whole range. The gaps between values also grow as values become larger. If particles are tracked using a global position, the accuracy of distance calculations will decrease as the position grows larger. The decreased accuracy can lead to spurious boundary crossings and change the results in undesirable ways. This can be particularly disastrous for optically thick problems where the distance traveled will become too small to resolve at larger positions and particles will not change position. In IMC simulations with implicit capture, those unmoving particles will still deposit energy in their cells leading to overheating. In the worst cases where the collision distance underflows to zero, the particles will fail to advance to the next census time and the simulation will stall unless a maximum particle iteration limit is applied. In contrast, the use of relative cell positions will ensure consistent distance calculations throughout the entirety of the problem space. Furthermore, relative positions have the advantage of allowing easy dynamic re-scaling of distances by only changing individual cell widths. This can be used to extend the range of cross-sections that can be accurately modeled in reduced-precision.

When developing a re-scaling approach it is important to first consider the units of the physical quantities involved. For reduced-precision it is generally preferable to scale values around unity. As an example the units used to represent distances could be re-scaled from cm to nm (or whatever units are most appropriate) for an optically thick problem to avoid underflow. Ensuring dimensional consistency in calculations that use values whose scale has been changed is essential. Challenges can arise when variables such as the wave-speed $c$ appear in multiple equations that have to be re-scaled in contradictory ways to preserve consistent units and remain in a representable range. One approach is to introduce extra scaling terms that multiply into specific equations as needed. This works especially well for intermediate calculations such as calculating the radiation energy sourced in a time-step. Further intermediate calculations can proceed with the scaled dimensions/magnitude and then the scaling terms can be divided out where convenient and the final results will have the correct units and scale. In cases where a scaling term cannot be divided out be sure to report the change to users. 

As an example, in half-precision the particle energy in sourcing has to be scaled to run with larger numbers of particles and smaller time-steps. Without scaling, the energy allotted to each particle would be too small to accurately represent when running with enough particles to reduce statistical noise to an acceptable degree. Scaling the particle energy by a constant factor of $100$ worked well for our problems, but this will vary. The scaled energies are used for the Monte Carlo simulation steps for more accurate energy deposition and are returned to their original scales before calculating the radiation and material energy densities at the end of each time-step. 

A more sophisticated re-scaling technique is required for quantities whose scale may vary over multiple orders of magnitude. This can often occur in transport problems with energy/frequency-dependent cross-sections.  In these cases there may not be a single constant scaling factor that can adequately cover the required range of values. Since interactions in Monte Carlo particle problems are typically governed by the product of the cross-section and distance traveled $\sigma * x$, as long as the scaling preserves this quantity the problem simulated should be the same. This allows us to model cross-sections that are too large to store in half-precision by scaling distances by a factor $S_F$ equal to the quotient of the larger cross-section divided by the representable one:

\begin{equation}
    \sigma_{large}*x = \sigma*\frac{\sigma_{large}}{\sigma}*x = \sigma*S_F*x
\end{equation}

If relative positions are used, the scaling factors can be dynamically changed when cross-sections change such as during scattering events or when a particle has reached the end of a time-step. Regarding time-steps, it is advisable to avoid using an absolute time variable to keep track of time-dependent occurrences in reduced-precision. Simulations should continue until a specified number of cycles (determined by time-step and desired maximum time) have completed rather than seeing if a time variable has reached the maximum time. When evaluating the current time of a particle within a simulation (for example when sourcing it at a given start time) it is also best to use a relative time between zero and the time-step length. Any particle that reaches the census to be passed along to a subsequent time-step should have its internal time reset to zero prior to the next simulation round.

\subsection{Floating-Point Arithmetic Manipulations}

 A common feature in Monte Carlo codes is a tallying step, where a large number of values must be summed. Care must be taken to prevent the accrual of significant round-off errors when computing tallies in reduced-precision arithmetic. These errors can be especially likely to occur when using a naive summation strategy where floating-point values are simply added sequentially. If the number of quantities $n$ to be added is small and they are all of a similar size, the error will be fairly small. However, if $n$ is large and/or the quantities vary greatly in magnitude, the error can become large with a worst case round-off error of order $\mathcal{O}(n)$ \cite{FPAccuracy93}. If a quantity to be added to a sum is smaller than the value stored in the last significant digit of the sum it will be rounded-off. This can be problematic for Monte Carlo tallies where many small values are added together. The tally may be accurate for values added early on, but round-off error will grow with the sum and a point may be reached where small additions to the tally make no contribution as they are rounded-off entirely.  

One summation method that can significantly reduce round-off errors is the pairwise summation method \cite{FPAccuracy93}. In pairwise summation, a list of numbers to be added is recursively partitioned into half-sized lists. The sub-lists are then added together in stages until the original list has been fully summed. The advantage of this method is that the partial sums to be added are more likely to be of the same order of magnitude, preventing the occurrence where smaller numbers are lost when added to a larger number. This method is most effective if the values in the original list are sorted by size prior to splitting to ensure each sub-list has similarly sized values at each stage of the summation. The worst case round-off error is of order $\mathcal{O}(\log n)$. When applied to a Monte Carlo tally of many small values, the halves added at each step of the summation process will be of a similar size and each of the small values will contribute to the final sum. 

An additional floating-point summation technique is Kahan summation \cite{KahanSum} which keeps a separate running compensation variable to collect small errors during summations. It has a worst case round-off of constant order $\mathcal{O}(1)$, at the expense of significantly more arithmetic operations for a given calculation \cite{FPAccuracy93}. From a practical standpoint, Kahan summation can be difficult to implement numerically for large tallies and thus pairwise summation was used for the results in this work. There are also methods known as Error-Free transformations which calculate the rounded result of a floating-point computation along with an accompanying rounding error \cite{OGITATakeshi2005Asad}. It may be possible to leverage this on a mixed-precision system to perform calculations in reduced-precision and then recover additional precision by converting the result and its rounding error into a higher-precision format.

Much greater consideration must be paid to the order of operations in reduced-precision floating point arithmetic to prevent the occurrence of over/underflow in calculations. Consider the multiplication of three floating-point numbers, two large and one small. If the two large numbers are multiplied together first, an overflow can occur and the calculation will be ruined. However, if one of the large numbers is multiplied by the small number first, the intermediate results will not overflow and can be safely multiplied by the second large number. Similar problems with underflow can occur when multiplying two small numbers and a large number together and can also be resolved by changing the order of operations. For thermal radiation transport problems, it can be advantageous to split the $T^4$ proportionality in the Stefan-Boltzmann law into two separate $T^2$ terms multiplied by the appropriate constant in the middle. The accuracy of results from more complicated expressions involving many small and large numbers can also be improved through similar strategies of reordering arithmetic operations. 

\subsection{Random Number Generation}
A crucial element of a good Monte Carlo code is a reliable random number generator. Many excellent generators have been developed; however, they typically work with higher-precision floating-point numbers. While the analysis of reduced-precision random number generation algorithms is beyond the scope of this work, a few practical suggestions are provided here. When generating half-precision random numbers it is preferable to use a direct generation method, rather than merely truncating a higher-precision floating-point random number into a reduced-precision format. This will help prevent bias in the statistics of the generated random numbers. In Monte Carlo applications, exponential distributions for particle collision distances are typically calculated as $d_{col} = \frac{-log(\xi)}{\sigma}$ using the inverse transform sampling method where $\xi$ is a uniform random variable $U(0,1)$. Using a purpose built random exponential distribution generator (randexp() in Julia) was found to produce collision distance results that more closely followed the expected distribution and avoided clumping in the tail of the distribution that could arise from directly taking the log of a reduced-precision $\xi$. Precautions should be taken when using reduced-precision variables in logarithmic, exponential, trigonometric, or sundry other special functions. In particular, check if the range of values produced by your random number generator is inclusive of zero and/or one as both these values can cause undesirable results such as infinities or NaNs when provided as input into certain functions. If those outputs are possible, a simple protection is to add an exception catcher that will re-roll for a new random number in cases that might otherwise cause the program to crash. Lastly, if you are running directly on reduced-precision hardware, it may be necessary to develop custom algorithms to accurately calculate various special functions if they are not natively included. 

\subsection{Batching}
Often the number of particles that are required in a Monte Carlo problem to obtain results that statistically converge to the desired degree will exceed the maximum 16-bit integer representable value of 65504. This limitation can be overcome by partitioning particles into smaller batches whose count does not exceed the maximum value. A simple partitioning strategy would evenly divide $N_{total}$ particles into $m$ batches of $N_{batch}$. Higher-order batching schemes can be developed if $N_{total}$ is so large that both $m$ and $N_{batch}$ will exceed their maximum representable values. These individual batches of particles can be executed with resulting tallies saved and combined at the end of the simulation, or each time-step for time-dependent problems. If best-practices regarding floating-point arithmetic are followed, the results from the batching method should be equivalent those obtained if the total number of particles were used without batching.

\section{Results} \label{sec:2}

The problem of thermal radiative transfer using the Implicit Monte Carlo (IMC) method was selected to test the efficacy of reduced-precision implementations. Thermal radiation problems can include many non-linear dependencies such as the $T^4$ black-body sourcing proportionality, complex frequency/temperature dependent cross-sections, and various heat capacities or equations of state for the material. Furthermore, many quantities in IMC can span orders of magnitude and there is a highly sensitive time-dependent coupling between the radiation transport and material energy update equations. If time-steps are too large, too much energy can be deposited in a cell leading to overheating and potential maximum principle violations \cite{40IMC}. Conversely, small time-steps (along with larger cell sizes) can lead to ``teleportation" errors where particles are non-physically emitted ahead of a wave-front at a given time. Thus, a delicate balance must be achieved between various numerical factors for accurate results. The numerical challenges for accurate modeling are exacerbated in reduced-precision and require the application of various scaling and floating-point arithmetic tricks for a successful implementation.

\subsection{Su \& Olson Benchmark IMC Problem}
This problem is an attempt to recreate the Su \& Olson Benchmark solutions for a non-equilibrium radiative transfer problem in an isotropically scattering medium \cite{SU19971035}. The problem features an initially cold, homogeneous, semi-infinite medium with an internal radiation source. The source has unit-strength over a width $x_0$ and remains active for a time $0 \leq \tau \leq \tau_0$. This problem features a material heat capacity with cubic temperature proportionality $c_v = \alpha T^3$  which allows for the radiation transport and material balance equations to be linearized and written in the following scaled form:

\begin{align}
    \big(\epsilon \frac{\partial}{\partial \tau} + \mu \frac{\partial}{\partial x} + 1\big) U(x,\mu,\tau) = \frac{c_a}{2} V(x,\tau) + \frac{c_s}{2} W(x,\tau) + Q(x,\mu,\tau), \\
    \frac{\partial V(x,\tau)}{\partial \tau} = c_a [W(x,\tau) - V(x,\tau)]
\end{align}

where 
\begin{align}
    U(x,\mu,\tau) = \frac{I(z,\mu,t)}{a T_{H}^4}, \quad V(x,\tau) = \Bigg[\frac{T(z,t)}{T_H}\Bigg]^4, \\
    W(x,\tau) = \int_{-1}^{1} d\mu U(x,\mu,\tau), \quad Q(x,\mu,\tau) = \frac{S(z,\mu,t)}{a T_{H}^4}.
\end{align}

The quantities $U$, $W$, $V$, and $Q$ are scaled radiation intensity, radiation energy density, material energy density, and radiation source, respectively. Additional variables include the absorbing and scattering ratios $c_a$ and $c_s$, as well as the scaled position and time $x$ and $\tau$. $T_H$ is a reference temperature which can be normalized along with the radiation constant $a$ and wave-speed $c$ to a value of $1$ in scaled units. The parameter $\epsilon = \frac{4 a}{\alpha}$ is also set to $1$.

Using an Implicit Monte Carlo treatment of this problem, the equations to solve become:

\begin{align}
    \frac{1}{c}\frac{\partial I}{\partial t} + \mu \frac{\partial I}{\partial x} + (\sigma_a + \sigma_s) I = \frac{f \sigma_a a c T^4}{2} + \frac{1}{2} \int_{-1}^{1} ((1-f) \sigma_a + \sigma_s) d\mu + \frac{Q_r}{2}, \\    
    \frac{\partial U_m}{\partial t} + f \sigma_a  a c T^4 = \int_{-1}^{1} f \sigma_a  I d\mu
\end{align}

where $f = \frac{1}{1+ \alpha \beta c \Delta t \sigma_a }$ is the Fleck Factor giving the effective absorption cross-section ratio.

Shown in Figures 1-4 are the material and radiation energy density results of double and half-precision IMC code implementations plotted against analytic benchmark values at different times for a problem with equal absorption and scattering ratios $c_a = c_s = 0.5$. The radiation source $Q_r$ is a unit source with width $x_0 = 0.5$ and an active duration of $\tau_0 = 10$. These problems were executed with 2000 time-steps of duration $\Delta t = 0.005$ or $\Delta t = 0.05$ to reach maximum times $\tau = 10$ or $\tau = 100$ respectively in dimensionless units. The cells have a scaled width $\Delta x = 0.01$ and the number of source particles introduced each time-step is $N_S = 500$, with particles apportioned according to the procedure described in the original Fleck \& Cummings IMC paper \cite{FLECK1971313}. The left-boundary at the origin is reflecting and there is a right vacuum boundary at $x = 10$ for the shorter simulation and $x = 20$ for the longer simulation. The vacuum boundary distances were chosen to be sufficiently far enough away from the origin for the IMC results to match the benchmark without having to simulate the whole half-space $0<x<\infty$. 

\begin{figure}[ht]
    \centering
    \includegraphics[width=0.6\textwidth]{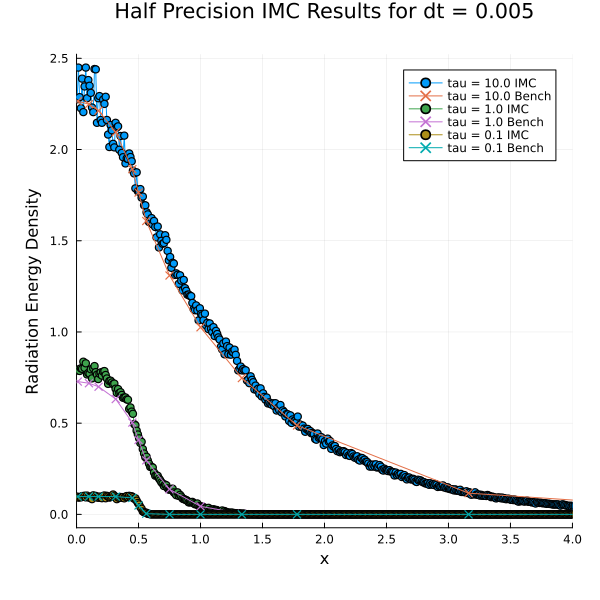}
    \caption{Half-Precision Radiation Energy Density Scatter Case}
    \label{fig:radenghalfscatter}
     \includegraphics[width=0.6\textwidth]{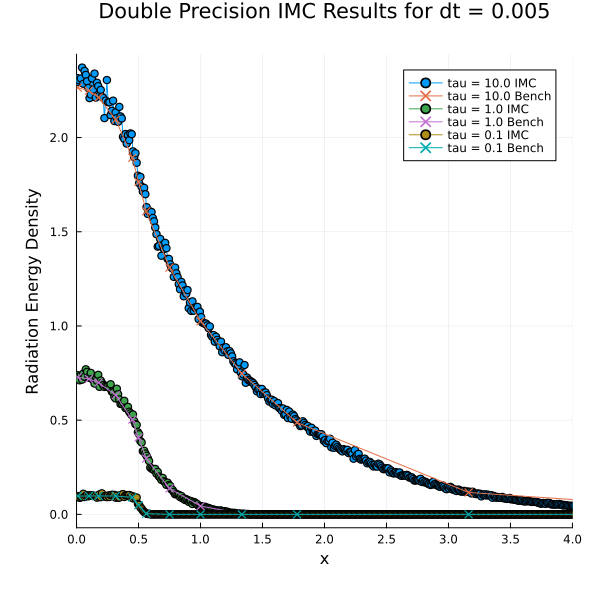}
    \caption{Double-Precision Radiation Energy Density Scatter Case}
    \label{fig:radengdoublescatter}
\end{figure}

\begin{figure}[ht]
    \centering
    \includegraphics[width=0.6\textwidth]{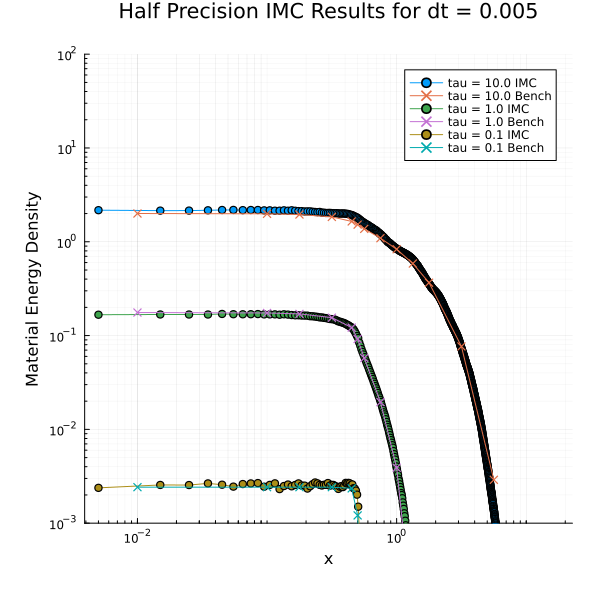}
    \caption{Half-Precision Material Energy Density Scatter Case}
    \label{fig:matenghalfscatter}
    \includegraphics[width=0.6\textwidth]{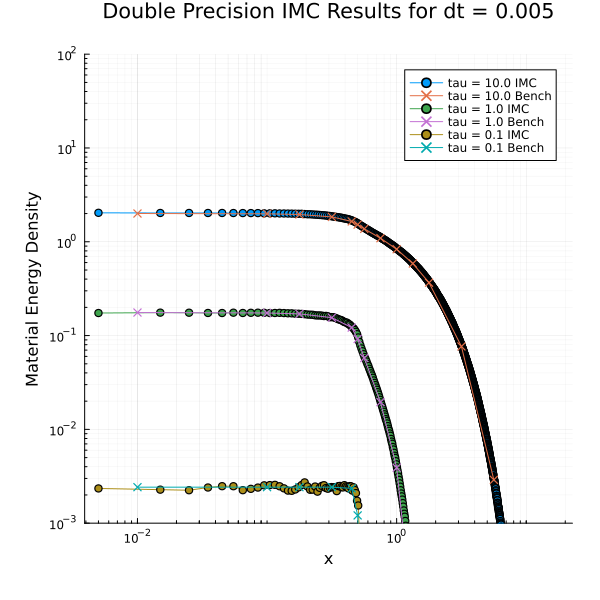}
    \caption{Double-Precision Material Energy Density Scatter Case}
    \label{fig:matengdoublescatter}
    
\end{figure}
 
Looking at the radiation energy density results, we can see the radiation energy density increases with time as more and more radiation energy is introduced into the system from the radiation source as well as from black-body radiation emitted by the material. The half and double-precision IMC results appear to converge to the benchmark solutions. The half-precision case appears to show a small amount of overheating in the region between the origin and $x_0$ containing the radiation source, but the overheating disappears outside of this region. The material energy densities for both cases match the benchmark values closely, with the half-precision case being only slightly more noisy. However, there is still good agreement with benchmark values in the tail towards the right of the slab.
The $L^2$ relative error norms for both precision cases calculated at points spaced at $0.05$ intervals between $0<x<4$ are tabulated in Table \ref{tab:l2error} below:
\begin{table}[ht]
    \centering
    \small 
    \caption{$L^2$ Relative Error Norms}
    \begin{tabularx}{\textwidth}{|c|c|c|c|c|}
        \hline
         \textbf{Time} & \textbf{Rad Energy Double} & \textbf{Rad Energy Half} & \textbf{Mat Energy Double} & \textbf{Mat Energy Half} \\ \hline
         \textbf{0.1} & $0.04275$ & $0.05606$ & $0.14890$ & $0.14702$ \\  \hline
         \textbf{1.0} & $0.03002$ & $0.10766$ & $0.01998$ & $0.04135$ \\  \hline
         \textbf{10.0} & $0.03460$ & $0.05101$ & $0.02038$ & $0.11021$ \\ \hline
    \end{tabularx}
    \label{tab:l2error}
\end{table}

We can observe that while some of the half-precision $L^2$ relative error norms are larger than their double-precision counterparts they are still reasonable and within the same order of magnitude. The norms for individual quantities are also generally consistent at different time-steps with only a couple outliers. The results from longer simulations up to $\tau = 100$ as well as those from pure absorber cases with no physical scattering can be found in the GitHub repository.

\section{Conclusions}
In this paper we described methodologies that can be used to implement an accurate Implicit Monte Carlo code in reduced-precision floating-point arithmetic. Scaling approaches, clever floating-point arithmetic algorithms, appropriate reduced-precision random number generators, and batching are all essential techniques for accurate reduced-precision IMC. Results from half and double-precision implementations of the Su Olson volume source problem were shown to agree well with the benchmark solution when using the aforementioned reduced-precision techniques. An ongoing limitation of reduced-precision implementations is the generation of less precise, albeit still accurate results. Future efforts can investigate approaches to recover additional precision from reduced-precision codes, dynamic re-scaling approaches to allow for a larger suite of problems to be accurately simulated, and the development of reduced-precision versions of more sophisticated IMC methods.

\printnomenclature 

\section*{Acknowledgements}
This work was supported by the U.S. Department of Energy through the Los Alamos National Laboratory. Los Alamos National Laboratory is operated by Triad National Security, LLC, for the National Nuclear Security Administration of U.S. Department of Energy (Contract No. 89233218CNA000001).

\bibliographystyle{IEEEtran}
\bibliography{refs}

\end{document}